\documentclass{appolb}
\begin{document}
%\date{\today}
\pagestyle{plain}
%% uncomment the following line to get equations numbered by (sec.num)
%\eqsec
\newcount\eLiNe\eLiNe=\inputlineno\advance\eLiNe by -1
\title{THE NEUTRINO FUTURE\\
  ---  COMMENTS HONORING GUSTAVO BRANCO  --- \footnote{To appear in the Proceedings of ``CP Violation and the Flavor Puzzle'', a Symposium held in Honor of Gustavo Branco in Lisbon, July 19-20, 2005.}
%
%\thanks{Send any remarks to {\tt acta@jetta.if.uj.edu.pl}}%
}
\author{Boris KAYSER
\address{Fermilab, MS 106, P.O. Box 500, Batavia IL 60510 USA} }
\maketitle

\begin{abstract}
We briefly report on a study intended to help shape the future of neutrino physics --- an area to which Professor Gustavo Branco has made distinguished contributions.
 \end{abstract}

%\section{Introduction}
Gustavo Branco's distinguished contributions include leading ones to neutrino physics, CP violation, and leptogenesis. To plan the future experimental exploration of some of the physics to which Prof. Branco has contributed, and which is near and dear to his heart, several studies focusing on neutrinos have been carried out. I would like to report on one of these, sponsored by four Divisions of the American Physical Society \cite{ref1}. This study's aim was to develop a coherent and effective strategy for the U.S. role in a global neutrino program. It was intended that the U.S. role complement and cooperate with the efforts in Europe and Asia \cite{ref2}.

We now have beautiful and compelling evidence that neutrinos have nonzero masses. This discovery has raised a number of interesting questions. The APS Multi-Divisional study has grouped these according to three themes. The first theme is ---  

\section*{NEUTRINOS AND THE NEW PARADIGM}

Even in the unlikely event that the neutrinos do not surprise us any more than they already have, we would like to answer the following questions:
\begin{itemize}
\item What are the masses of the neutrinos? Does the neutrino mass spectrum resemble the spectra of the quarks and the charged leptons, or is it an inverted version of those other spectra? How high above the zero of mass does the entire spectrum lie?
\item What is the pattern of mixing among the different types of neutrinos? How large is $\theta_{13}$, the small mixing angle on whose size the magnitude of CP violation in neutrino oscillation depends? Is the very large atmospheric mixing angle $\theta_{23}$ maximal (in which case an underlying symmetry may well be involved)?
\item Are neutrinos their own antiparticles? If they are, that would make them different from all the other fermionic constituents of matter.
\item Do neutrinos violate the symmetry CP? In particular, do corresponding neutrino and antineutrino oscillation probabilities differ?
\end{itemize}

The second theme is --- 

\section*{NEUTRINOS AND THE UNEXPECTED}

The study of neutrinos has been marked by surprises. The masses of the neutrinos have proved to be very much smaller than those of any other known particles. Two of the neutrino mixing angles have turned out to be very large, while all of the quark mixing angles are small. The low value of the solar $\nu_e$ flux arriving at earth, compared to the predicted rate of neutrino production by the sun, was a surprise. So was the low value of the $\nu_\mu / \nu_e$ ratio in the atmospheric neutrino flux arriving at underground detectors. It would be surprising if further surprises were not in store. The open questions include --- 
\begin{itemize}
\item Are there ``sterile'' neutrinos --- neutrinos that do not participate in the Standard Model weak interactions, or any other known interactions except gravity? Correspondingly, are there more than  three neutrino mass eigenstates? \item Do neutrinos have unexpected or exotic properties, such as magnetic dipole moments that are many orders of magnitude larger than those predicted by the Standard Model and large enough to be observed?
\item What do neutrinos have to tell us about the intriguing proposals for new models of fundamental physics? The See-Saw Mechanism relates neutrino masses to physics at the high-mass scale where the weak, electromagnetic, and strong interactions appear to become unified. What can we learn about Grand Unified Theories by studying the neutrinos?
\end{itemize}

The third theme is --- 

\section*{NEUTRINOS AND THE COSMOS}

Neutrinos and photons are by far the most abundant particles in the universe. Neutrinos have played an important role in shaping the large-scale structure of the universe, and in determining its character. We would like to know --- 
\begin{itemize}
\item What, precisely, has been the role of neutrinos in shaping the universe? What can we learn about the absolute scale of neutrino mass from cosmological observations?
\item Is CP violation by neutrinos the key to understanding the matter Ð antimatter asymmetry of the universe?
\item What can neutrinos, acting as messengers, reveal about the deep interior of the earth and sun, about supernovae, and about ultra-high-energy astrophysical phenomena whose produced photons cannot reach us?
\end{itemize}

The APS study urged that two future experimental programs be given high priority. To wit:

\vspace{.2in}
{\em We recommend, as a high priority, that a phased program of (increasingly) sensitive searches for neutrinoless nuclear double beta decay be initiated as soon as possible.}
\vspace{.2in}

Neutrinoless nuclear double beta decay ($0\nu\beta\beta$ decay) is the process in which one nucleus decays into another plus two electrons, and nothing else. Observation of this process at any level would establish that --- 

\begin{itemize}
\item The lepton number L that distinguishes between leptons and antileptons is not conserved.
\item Neutrinos are Majorana particles. That is, each neutrino of definite mass is its own antiparticle.
\item Nature (although not the Standard Model) contains Majorana neutrino masses.
\item The origin of neutrino mass involves physics different from that which gives masses to the charged leptons, quarks, nucleons, humans, the earth, and galaxies.
\end{itemize}

Thus, observation of $0\nu\beta\beta$ decay would establish that neutrinos and their masses are very distinctive indeed. There is interest in several areas of the world in searching for this decay. It is hoped that U.S. scientists, some of whom have very high relevant expertise, can make a major contribution to the global effort.
  
\vspace{.2in}
{\em We recommend, as a high priority, a comprehensive U.S. program to complete our understanding of neutrino mixing, to determine the character of the neutrino mass spectrum, and to search for CP violation among neutrinos.}
\vspace{.2in}
  
	This program should have three components:
  
\vspace{.1in}
A.	An expeditiously-deployed reactor experiment with sensitivity to the small mixing angle $\theta_{13}$ down to $\sin^2 2\theta_{13} = 0.01$. 

B.	A timely accelerator experiment with $\theta_{13}$ sensitivity comparable to that of the reactor experiment, and with sensitivity, through matter effects, to whether the neutrino mass spectrum is normal (i.e., quark-like) or inverted.

C.	A megawatt-class proton driver and neutrino superbeam with an appropriate very large detector capable of observing CP violation.
 \vspace{.1in}

	Completion of our understanding of mixing will focus on $\theta_{13}$, with attention to $\theta_{23}$ as well.  Determining the approximate size of $\theta_{13}$ will discriminate between many models \cite{ref3}. In some models, $\sin\theta_{13}$ is naturally of order $\Delta m^2_{\mathrm{sol}} / \Delta m^2_{\mathrm{atm}} \cong 1/30$, where $\Delta m^2_{\mathrm{sol}}$ and $\Delta m^2_{\mathrm{atm}}$  are the solar and atmospheric squared-mass splittings, respectively. In these models, $\sin^2 2\theta_{13}  \sim 0.004$. In other models, $\sin \theta_{13}$ is of order $(\Delta m^2_{\mathrm{sol}} /\Delta m^2_ {\mathrm{atm}})^{1/2} \cong 1/6$, so that $\sin^2 2\theta_{13} \sim 0.1$, very close to the present upper bound. Thus, learning whether $\sin^2 2\theta_{13}$ is larger or smaller than, say, 0.01 will discriminate between these two classes of models. In addition, if $\sin^2 2\theta_{13}$ proves to be less than 0.01, then we will have learned that a neutrino factory, producing neutrinos through the decays of muons in a storage ring, or a beta beam, producing them in the decays of radioactive ions, will be necessary to observe CP violation and to determine whether the neutrino mass spectrum is normal or inverted. But if $\sin^2 2\theta_{13} > (0.01- 0.02)$, then both of these issues can be addressed via conventional, albeit high intensity, neutrino beams.

	An experiment seeking to learn about $\theta_{13}$ by studying the disappearance of some of the   $\overline{\nu_e}$ flux from a reactor has the advantage that this disappearance will not depend on any of the other unknown neutrino properties or parameters. Consequently, such an experiment can determine $\theta_{13}$ cleanly. In contrast, experiments with accelerator neutrinos can access all the neutrino properties we wish to study: $\theta_{13}$ and $\theta_{23}$, the normal or inverted character of the mass spectrum, and CP violation. However, in the accelerator experiments, various neutrino parameters are intertwined in the quantities that one actually measures, so a clean determination of $\theta_{13}$ will be very useful to the overall program. 

	The atmospheric mixing angle $\theta_{23}$ mixes $\nu_\mu$ and $\nu_\tau$. If it turns out that $\theta_{23}$ is not maximal (45$^\circ$), then we will want to know whether it lies below or above 45$^\circ$. In combination with other information, this will tell us whether the heaviest neutrino mass eigenstate is more $\nu_\tau$ than $\nu_\mu$, as naively expected, or the other way around.

	Turning to the character of the mass spectrum, we note that, generically, Grand Unified Theories (GUTS) favor a normal neutrino mass spectrum --- one resembling the charged-lepton and quark spectra. The reason is simply that in GUTS, the neutrinos, charged leptons and quarks are all related, so we expect their spectra to be similar. However, some classes of string theories lead one to anticipate an inverted neutrino mass spectrum. Thus, the actual nature of this spectrum is an interesting question.

	Of course, if, as suggested by the LSND experiment, there are more than three neutrino mass eigenstates, then the neutrino mass spectrum is {\em very} different from the charged lepton and quark spectra: the neutrino spectrum has more states than its charged lepton and quark counterparts. Such a finding would require careful rethinking of the future neutrino program.

	Assuming that there are just three mass eigenstates, the (mass)$^2$ spectrum consists of a closely-spaced pair of states, $\nu_1$ and $\nu_2$, separated from each other by the solar (mass)$^2$ splitting, $\Delta m^2_{\mathrm{sol}} \cong 8.0 \times 10^{-5}$ eV$^2$ \cite{ref4}, and a third state, $\nu_3$, separated from the $\nu_1 - \nu_2$ pair by the thirty-fold larger atmospheric (mass)$^2$ splitting, $\Delta m^2_{\mathrm{atm}} \cong 2.4 \times 10^{-3}$ eV$^2$ \cite{ref5} . If $\nu_3$ lies above the $\nu_1 - \nu_2$ pair, then the (mass)$^2$ spectrum is normal, while if it lies below the $\nu_1 - \nu_2$ pair, then the spectrum is inverted. Provided that $\theta_{13}$ is not too tiny \cite{ref6}, we can determine whether the spectrum is normal or inverted by exploiting the matter effect. 
This leads to an asymmetry between $\Delta m^2_{\mathrm{atm}}$-driven $\nu_\mu \rightarrow \nu_\e$ and $\overline{\nu_\mu} \rightarrow \overline{\nu_e}$ oscillations, and the sign of this asymmetry depends on whether the spectrum is normal or inverted. To be sensitive to the $\Delta m^2_{\mathrm{atm}}$-driven oscillations, an experiment needs to place its far detector near the first maximum of these oscillations. This maximum occurs at $L/E \cong$ 500 km/GeV, where $L$ is the baseline of the experiment, and $E$ is the neutrino energy.  Since the matter effect grows with energy, it is preferable to work with higher $E$, which requires one to have a correspondingly larger $L$. A combination such as $E \sim$ 2 GeV and $L \sim$ 800 km, which is possible in the U.S., could make a key contribution to the global study of neutrinos by enabling us to determine the character --- normal or inverted --- of the spectrum.

	Like the matter effect, CP violation also will lead to an asymmetry between neutrino and antineutrino oscillations. One will have to disentangle the spectrum-dependent matter effect from genuine CP violation in the asymmetry that is actually observed. To this end, the proposed Japanese and U.S. accelerator neutrino programs, which will operate at energies differing from each other by a factor of three and consequently will involve matter effects of quite different size, can play strongly complementary roles. 

	A megawatt-class proton driver, or equivalent high-intensity neutrino source, and a suitable large detector will be needed if we are to be able to establish the presence of CP violation for any value of $\sin^2 2\theta_{13}$ above (0.01 -- 0.02). 

The observation of CP violation in neutrino oscillation would be very interesting not only because it would establish that CP violation is not a peculiarity of quarks, but also because of the possible connection to the matter-antimatter asymmetry of the universe. In the See-Saw Mechanism, the most popular explanation of the incredible lightness of neutrinos, the light neutrinos $\nu$ are accompanied by very heavy ``see-saw partner'' neutrinos N. These heavy neutrinos are much too heavy to be made in laboratory experiments of today, but would have been made in the hot Big Bang. Subsequently, each of them would have decayed into a final state that includes an ordinary lepton or antilepton. 

	If oscillation of today's light neutrinos violates CP, then, quite likely, so does the decay of the heavy see-saw partners of these neutrinos. Thus, the decays of these heavy neutrinos in the early universe would have produced a world with unequal numbers of leptons and antileptons. Non-perturbative Standard Model sphaleron processes would then have reprocessed this lepton-antilepton asymmetry into one that involves both baryons and leptons, producing the baryonic and leptonic matter-antimatter asymmetry that we see today. This possibility makes it very interesting indeed to see whether light neutrino oscillation does indeed violate CP.

	The production of the matter-antimatter asymmetry through the decays of heavy neutrinos is known as leptogenesis \cite{ref7}.  Gustavo Branco and his collaborators have made very important contributions to our understanding of this hypothesis. In particular, they have illuminated the connections between leptogenesis and the quantities that we can study in laboratory experiments of today. 

	Congratulations, Gustavo, for all you have done {\em so far}, and thank you in advance for all you {\em will} do in the future! 

\section*{Acknowledgments}

The results of the APS Multi-Divisional study grew out of the dedicated efforts of many people. It is a pleasure to thank my colleagues in this study for their insights, creativity, wisdom, cooperative spirit, and energy.

It is also a pleasure to thank Gustavo Branco, Jo\~{a}o Silva, and Gui Rebelo for truly exceptional hospitality during a three-week visit to Lisbon.

\end{document}